\documentclass[final,5p,times,twocolumn,authoryear]{elsarticle}

\usepackage{url}
\usepackage{mathtools,amssymb,amsfonts}
\usepackage{graphicx}
\usepackage{microtype}
\usepackage{lipsum}
\usepackage{lineno}
\usepackage{blindtext}
\usepackage{natbib}
\usepackage{textcomp}
\usepackage{amsmath}
\usepackage{booktabs}
\usepackage{multirow}
\usepackage{tikz, pgfplots}
    \pgfplotsset{compat=1.15}
\usepackage{color}
\usepackage{siunitx}
\usepackage{todonotes}

\newcommand{\bs}[1]{\boldsymbol{#1}}

\begin{document}
\begin{frontmatter}

\title{Vehicle single track modeling using physics guided neural differential
equations \tnoteref{t1}}

\tnotetext[t1]{Supplementary code is available at:~\citep{hybrid_node2022}}

\author[1]{Stephan~Rhode\corref{cor1}}
\ead{stephan.rhode@bosch.com}
\cortext[cor1]{Corresponding author}

\author[1]{Fabian Jarmolowitz}
\ead{fabian.jarmolowitz@bosch.com}

\author[1]{Felix Berkel}
\ead{felix.berkel@bosch.com}

\affiliation[1]{organization={Bosch Center for Artificial Intelligence,
            Robert Bosch GmbH},
            city={Renningen},
            country={Germany}}

\begin{abstract}
In this paper, we follow the physics guided modeling approach and integrate
a neural differential equation network into the physical structure of a
vehicle single track model. By relying on the kinematic relations of the single
track ordinary differential equations (ODE), a small neural network and few
training samples are sufficient to substantially improve the model accuracy
compared with a pure physics based vehicle single track model. To be more
precise, the sum of squared error is reduced by 68\% in the considered
scenario. In addition, it is demonstrated that the prediction capabilities of
the physics guided neural ODE model are superior compared with a pure black
box neural differential equation approach.
\end{abstract}

%


\end{frontmatter}

\section{Introduction}\label{sec:intro}
Vehicle dynamics models are an essential part of advanced planning, control,
and state estimation algorithms of autonomous vehicles. Thus, accurate
models of the vehicle dynamics play a vital role to guarantee safe and robust
vehicle operation. Add to the accuracy requirement, fast computation of the
vehicle dynamics models is desired to meet real time constraints in vehicle
motion control applications. Currently, most vehicle dynamics models are
built from physical principles and comprise vehicle specific parameters. These
parameters determine the model accuracy significantly, which is treated by
parameter optimization in many practical applications today. On the other
hand, the predefined structure of the physical vehicle models may cause
inaccuracy due to missing physical effects. Hence, in many cases it is not
possible to meet the corresponding accuracy and real-time requirements with
pure physics based models. With advanced sensing and upcoming
connectivity via vehicle-to-everything (V2X), the availability of vehicle
dynamics data is consistently increasing. As a consequence, there is much
potential to exploit this data for the model improvement by machine learning.

The modelling of dynamic processes like vehicle dynamics, are facing a
long-standing research interest. As a structured and very compact form,
ordinary differential equations (ODE) are the dominating mathematical
framework to describe these dynamic processes. The inference of ODEs from
data is known as system identification~\citep{soderstrom1989system,ljung_system_1999},
which is nowadays heavily influenced by machine learning~\citep{schoukens_2019}.
The challenge in system identification is to design and infer a model that
catches the important dynamics without over-fit. One option is to use the
laws of physics as building blocks for the ODE model. These models are called
white box models. The extension of white box models are grey box models,
where some model parameters are optimized on given measurement data of the
true system. On the contrary, black box models specify the right hand side of
the ODE with rather generic ansatz functions like polynomials or neural
networks, which are learned from informative measurement data of the true
system~\citep{schoukens_2019}. Moreover, researchers combined physics based
modeling and machine learning in the class of physics guided modeling, and
we will focus on the subclass called hybrid modeling in the
sequel.

Except for white box models, all other modeling methods contain unknown
parameters, which have to be learned from time series data of the
true system. The more information in structure or via regularization is
specified a priori, the fewer data is needed to achieve high model accuracy.
Therefore, physics guided models are promising, because they offer the
possibility to effortlessly integrate a priori information derived from
physics in machine learning frame works.

\subsection{Related literature}\label{subsec:related-literature}
We incorporate the neural differential equation (NODE)
approach~\citep{Chen2018} as black box method to derive a dynamical
vehicle single track model. Add to this, we combine NODEs with
differential equations of a known vehicle single track model to derive a
hybrid NODE model and compare the black box NODE and the hybrid NODE model
with a state-of-the-art ODE single track model. Accordingly, we will discuss
physics guided and hybrid vehicle models from the literature at first and
proceed with black box, white box, and state estimation topic in the following.

A comparative study with NODEs as black box approach for several applications
including vehicle dynamics is presented in~\citep{Rahman2022}. A superior
accuracy of NODEs is concluded compared to discrete-time linear and state-space
neural models. The formulation in continuous-time of NODEs in combination
with adaptive step-size integration schemes is suspected to be the cause for
the high accuracy of the NODEs.

\citet{Thummerer2022} applied a hybrid NODE model to predict the consumption
of an electric vehicle for a standard drive cycle. The model considers
pure longitudinal dynamics and showed superior accuracy compared with a
baseline physical model, although only a single drive cycle was used as
training set. This emphasizes that the physics guided modeling techniques
require few training data to find accurate predictions. \citet{Graeber2019}
present a gated recurrent neural network and a hybrid neural network
to estimate the vehicle sideslip angle. The combination of the physics based
kinematic model and the recurrent neural network showed competitive results
compared with an Unscented Kalman Filter approach, which functioned as
state-of-the-art benchmark. \citet{Sieberg2022} secure neural network
predictions of the vehicle roll angle by fusing predictions of a parallel
physical model through an Unscented Kalman Filter. The level of fusion is
controlled by a reliability estimate of the neural network. Hence, the
presented architecture requires to run two models in parallel and resembles
the general idea of interacting multiple model estimation~\citep{Mazor1998}.

\citet{James2020} present a comparison of different approaches for modeling
pure longitudinal vehicle dynamics including a neural network approach
derived from physics. The neural network is trained in a collocation-based
fashion, in which the derivatives are estimated using a smoothing Gaussian
function. In summary the neural network performed best, although a linear
model was favored due to its simplicity while having comparable accuracy in
the considered driving scenarios.

In~\citep{Hermansdorfer_2020} an end-to-end trained neural network with gated
recurrent layers was used in a black box approach to reproduce vehicle
longitudinal and lateral dynamics. The network showed better accuracy than
the considered physics based model. Other combined longitudinal and lateral
vehicle models based on neural networks were shown  in~\citep{Devineau2018,
    Yim2004,Spielberg2019}. In addition, the vehicle sideslip angle, which is
important for electronic stability program (ESP), was modeled by neural networks
in~\citep{Bonfitto2020, Melzi2011} and~\citep{Essa2021} provide a performance
assessment of different neural network architectures for sideslip angle
estimation. Specifically, Feed-Forward Neural Networks, Recurrent Neural
Networks, Long Short-Term Memory units, and Gated Recurrent Units were
compared with respect to accuracy, error variance, and computational effort
by training time and estimation time. The Feed-Forward Neural Networks
achieved higher accuracy than the other networks with recurrence, but the
Gated Recurrent Units passed the Feed-Forward networks in training time.

Most research in vehicle dynamics modeling has been conducted in the
field of white box modeling for vehicle state prediction
and estimation. The optimal control algorithms for longitudinal and lateral control
in~\citep{Katriniok2013, Liniger_2014} utilize white box vehicle
kinematics and tire models to predict vehicle states like acceleration, yaw
rate, and wheel steering angle in a model predictive control application.
Moreover,~\citep{Yi_2016} introduced an elaborate white box vehicle
dynamics model with seven states in vehicle trajectory planning
with model predictive control for critical maneuvers. In vehicle state
estimation, physics based white box models are used together with
sensor information to fuse uncertain model predictions and measurements with
help of probabilistic adaptive filters, see the numerous
applications referenced in~\citep{Kanwar2019, Guo2018}. Add to this, the
combination of vehicle state estimation by adaptive filtering and parameter
estimation of vehicle grey box models is known as dual
filtering~\citep{Wenzel2006}, which requires two parallel running Kalman
filters to estimate vehicle states and parameters simultaneously.

There is a growing body of application-oriented research with physics guided
hybrid modeling outside of vehicle dynamics modeling. \citet{Roehrl2020}
apply physics informed neural ordinary differential equations to derive a
hybrid model of a cart pole. Known physics was modeled via lagrangian
mechanics with certain parameters of gravity, lengths, masses, and
moment of inertia. Uncertain friction of cart and pole was adjusted by the
involved neural network, which contained two layers of fifty neurons each.
This hybrid model showed superior accuracy compared with a pure black box
model and a white box ODE model. \citet{Viana2021} design recurrent neural
networks for numerical integration of ordinary differential equations as
directed graph. By this, the nodes in the graph are treated as
physics-informed kernels. Extra data driven nodes were added to the graph to
model the missing physics via training. This method was tested on fatigue of
aircraft fuselage panels, corrosion-fatigue in aircraft wings, and wind
turbine bearing fatigue examples.

\subsection{Contribution}\label{subsec:contribution}

We investigate three modeling approaches, namely a simplified white-box
model (ODE), a pure back-box (neural ODE), and a hybrid model (UDE), for
combined lateral and longitudinal vehicle dynamics.
The simplified white-box model cannot reproduce the behavior of the
reference system in dynamic situations. The pure black-box model shows
improved performance by a reduction of the sum of squared error, i.e., the
error between prediction of the trained model and the actual measurements,
by 63\% compared to error of the white-box model in the considered scenario.
However, it requires many data points for training which diminishes the
practical suitability. The hybrid model, which combines the kinematic part
from the white-box model and a black-box model for the dynamic equations,
shows superior performance compared with the white-box model by a reduction
of the sum of squared error by 68\% in the considered scenario.
In addition, the hybrid model shows improved performance compared to the
black-box model and requires significantly less training points than the
black-box model which makes it appealing for practical usage.
To our best knowledge, these investigations for combined lateral and
longitudinal vehicle dynamics represent a novelty in the literature.
Moreover, the programming code in \texttt{Julia} is available as
supplementary material for the reader~\citep{hybrid_node2022}.

\subsection{Outline of the paper}\label{subsec:outline-of-the-paper}

This work is organized as follows.
Section~\ref{sec:physics_guided_learning} reviews modeling
methods from physics guided machine learning and introduces neural
differential equations and universal differential equations.
Section~\ref{sec:experiments} explains conducted experiments, applied model
structures, generation of training and validation data, and how the training
was processed. Section~\ref{sec:results_discussion} provides results for each
model and the result discussion, while Section~\ref{sec:conclusions}
concludes this paper.

\section{Physics guided machine
learning}\label{sec:physics_guided_learning}
There is growing consensus that complex engineering problems require modeling
techniques which combine predictability, generalization and interpretability on
the one hand and adaptation to data on the other
hand~\citep{Karpatne2017,Rai2020,Rueden2021}. While physical modeling usually
yields good interpretable models on high generalization, classical machine
learning methods like neural networks outperform physical modeling with
respect to adaptation to training data. However, this adaptation comes with
the price of large required training samples to achieve adequate model
accuracy. In classical machine learning, large training data sets are required
to train neural networks towards physical consistency and accurate
predictions in out of sample scenarios. However, data acquisition campaigns
deliver limited observation data in engineering problems. Specifically, the
effort to generate data from vehicle test drives is high and will remain a
cost factor. Therefore, researchers found techniques to combine
predictability, interpretability, and sample efficiency of physical modeling
with universal approximation of machine learning.

\citet{Willard2020} provide a taxonomy for physics guided machine learning.
There are four groups of methods; (i) physics guided loss function, (ii)
physics guided initialization, (iii) physics guided design of architecture,
and (iv) hybrid modeling. One example of physics guided loss function group
are physics-informed neural networks~\citep{Raissi2019} where the cost
function becomes a sum of supervised training error of the neural network and
physics based costs, which can be defined as physical laws of conservation
for instance. The second group physics guided initialization comprise methods
which consider physical laws or contextual knowledge in initialization of
neural network weights prior training. Compared with standard random
initialization, physics guided initialization results in accelerated training
on fewer training samples. In addition, local minima can be avoided with
physics guided initialization. Add to physics guided cost functions and
initialization, physics guided design of architecture encodes physical
consistency or other physics inspired structure in neural networks. One
example of this third group are neural differential equations (neural ODE)
~\citep{Chen2018}, which resemble the structure of differential equations and
will be introduced in Section~\ref{subsec:neural-differential-equations}.

Note that physics guided loss function, initialization, and design of
architecture focus on augmenting existing machine learning methods with
physical knowledge. Instead, the fourth group hybrid modeling combines
physical models with machine learning models, which means that both model
types operate simultaneously. In hybrid modeling, there are numerous
structures available from simple residual modeling, where a machine learning
model reduces imperfection of physical model, to structures where parts of the
physical model are replaced by a neural network. How to combine physics and
neural networks strongly depends on the domain and the knowledge about this
domain. In addition, the known shortcomings of the physical model influence
the way of combining physical model and network. Therefore, the combination
can be seen as a modeling task where a hybrid modeling paradigm is used.
This combined structure is frequently called universal differential equation
(UDE)~\citep{Rackauckas2020} and discussed in
Section~\ref{subsec:physics-informed-neural-differential-equations}.

\subsection{Neural differential equations}
\label{subsec:neural-differential-equations}

Neural ordinary differential equations (neural ODE) are a system of
differential equations specified by a neural network~\citep{Chen2018}.

\begin{equation}
    \frac{\mathrm{d}\bs{x}}{\mathrm{d}t} = \mathrm{NN}(\bs{x}, \bs{u},
    \bs{\theta}, t),
    \label{eq:general_node}
\end{equation}
where $\bs{x}$ are the states, $\bs{u}$ is a vector of exogenous
inputs, $\bs{\theta}$ contains the neural network weights, and $t$ is the
time.

The ODE right hand side describes the dynamical evolution in time of the system
states. The inference of the right hand side is known as system identification and 
leads to compact models describing a learned dynamic
behaviour~\citep{ljung_system_1999}.

Assume a scalar loss function of the current state $\bs{x}(t_1)$

\begin{align}
    \mathcal{L}(\bs{x}(t_1)) &= \mathcal{L}\left( \bs{x}(t_0)+\int_{t_0}^{t_1}\mathrm{NN}
    (\bs{x}, \bs{u},
    \bs{\theta}, t)dt \right)\\
    &=\mathcal{L}(\text{ODESolve}(\bs{x}(t_0), \mathrm{NN}, \bs{u},
    \bs{\theta}, t_0,t_1)),
    \label{eq:loss_function}
\end{align}

where the loss function contains a numeric integration scheme
$\text{ODESolve}$. The integration scheme is required because an analytic
solution of an arbitrary ODE is not known in general. Thus, the gradient of
the loss $d\mathcal{L}/d\bs{\theta}$ has to be propagated through the ODE
solver.

The calculation of the gradient of the solver with respect to states and
parameters can be problematic, especially for solvers with adaptive step
length. Therefore, the neural ODE method in~\citep{Chen2018} proposes adjoint
sensitivity analysis~\citep{Pontryagin_1962} for back-propagation of the
loss function's gradient over time with respect to the initial state $\bs{x}
(t_0)$.

The adjoint ODE system is defined as

\begin{equation}
    \frac{d\bs{a}(t)}{dt}=-\bs{a}(t)^T \frac{\delta \mathrm{NN}(\bs{x},
    \bs{u}, \bs{\theta}, t),}{\delta \bs{x}(t)}
    \label{eq:adjoint_system}
\end{equation}

with the adjoint variable $\bs{a}(t)=\delta \mathcal{L} /\delta \bs{x}(t)$.
By using the adjoint, the gradient of the loss can be formulated as

\begin{equation}
    \frac{\mathrm{d} \mathcal{L}}{\mathrm{d} \bs{\theta}} = -
    \int_{t_1}^{t_0}\bs{a}(t)^T
    \frac{\delta \mathrm{NN}(\bs{x}, \bs{u}, \bs{\theta}, t),}{\delta
    \bs{\theta}} dt.
    \label{eq:loss_gradient}
\end{equation}

Equations~\eqref{eq:general_node},~\eqref{eq:adjoint_system},
~\eqref{eq:loss_gradient} can be integrated backwards from $t_1$ to $t_0$
within a single call of an ODE solver by defining the initial augmented state as
$\bs{x}_{\mathrm{aug}}(t_1)=[\bs{x}(t_1), \delta \mathcal{L} / \delta \bs{x}
(t_1), \bs{0}_{|\theta|}]$. Numerical integration yields the final state
$\bs{x}_{\mathrm{aug}}(t_0)=[\bs{x}(t_0), \delta \mathcal{L} / \delta \bs{x}(t_0), \delta \mathcal{L} / \delta
\theta]$ provide the necessary gradients for training. Due to the use of
the adjoint method, this approach does not need any gradient depending on
the ODE solver.

Nevertheless, using the adjoint system is not the only option to calculate
the gradient of the loss function. It is also possible to compute the
gradient via a discretize-then-optimize approach by applying automatic
differentiation (AD) directly on the combined system of ODE solver and ODE
system. This can be done via forward mode or reverse mode automatic
differentiation. The forward mode is a viable option for systems with fewer
parameters~\citep{Ma_2021}.

\subsection{Physics guided neural differential
equations}\label{subsec:physics-informed-neural-differential-equations}

The integration of physical laws into a neural ODEs is a technique from
physics guided machine learning, which is known as hybrid modeling and
defined as universal differential equation (UDE) in~\citep{Rackauckas2020}.

\begin{equation}
    \frac{\mathrm{d}\bs{x}}{\mathrm{d}t} = f(\bs{x}, \bs{u}, \mathrm{NN}
    (\bs{x}, \bs{u}, \bs{\theta}, t), t)
    \label{eq:ude}
\end{equation}

The UDE form allows to model parts of the dynamical states by first
principles and other parts by a neural network. Add to this split between
the first principle state equations and neural ODE, an UDE can
also be used as correction term when the entire state vector is initially
computed by physical laws to enhance accuracy of the physical differential
equation. Another way of coupling physical models with neural ODEs is
discussed in~\citep{Thummerer2022}, where the physical model was exported
from Modelica tool into a Functional Mock-up Unit (FMU) and coupled with a
neural ODE in \texttt{Julia} programming language. \citet{Thummerer2022} call
this model structure hybrid neural ODE, which can be seen as synonym for
universal differential equation.

\section{Experiments}\label{sec:experiments}
In this section three model structures were compared in their accuracy and complexity by running experiments on data generated with a complex reference model derived from first principles.
The first model, called ODE in the following, is derived from linearization of the complex reference model. 
The second and third models are a pure data driven neural ODE model and a
physics guided UDE model.
The latter merges first principles and data driven modeling.

\begin{figure}
	\centering
	\includegraphics[width=0.9\columnwidth]{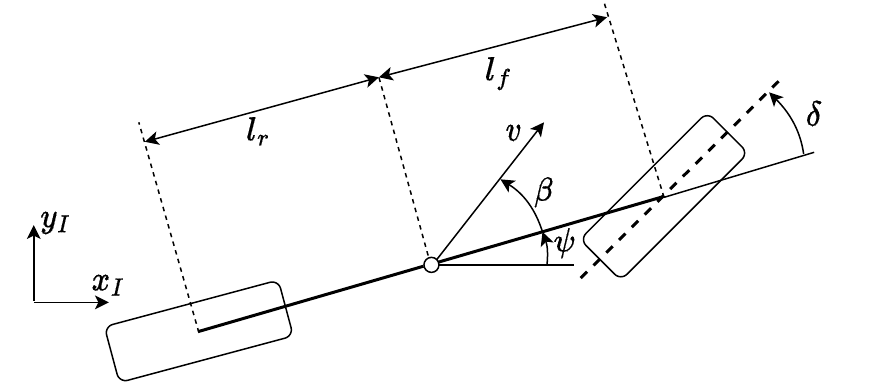}
	\caption{Top view of vehicle single track and single track drift model.
		The dot in center of the figure represents the center of gravity.}
	\label{fig:single_track_model}
\end{figure}

All considered models belong to the class of single track models. 
Single track models are one representation of vehicle dynamics models where
the front and rear tires are lumped together.
Figure~\ref{fig:single_track_model} shows the top view of a general single
track model with the global coordinate system ($x_I, y_I$), angles, and
velocity vector. $l_f, l_r$ are the front and rear distance from axle to the
vehicle's center of gravity.
These models are known to represent vehicle dynamics with high
accuracy upon medium longitudinal and lateral dynamics, see~\citep{Schramm2018} for more information. 
If high dynamics are of interest, more complex multi-body models can be used.

\subsection{Reference model}\label{subsec:reference-model}
Training and validation data was generated by simulation of the vehicle single track drift model of Common  Road's\footnote{https://commonroad.in.tum.de/}
\texttt{Python} library~\citep{CommonRoad}. 
Its state vector
\begin{equation}
\bs{x} = [\Delta x, \Delta y, \psi, \delta, v, \beta, \omega, \omega_{\text{f}}, \omega_{\text{r}}]^{\top}
\label{eq:state_drift}
\end{equation}
consists of the position in x and y ($\Delta x, \Delta y$), the yaw angle
$\psi$, the steer angle $\delta$, velocity $v$, slip angle $\beta$, the
yaw rate $\omega$, as well as, the front and rear wheel speed $\omega_{\text{f}}$ and $\omega_{\text{r}}$. The exogenous input
\begin{equation}
\bs{u}=[a_x, v_\delta]^\top
\label{eq:ex-input}
\end{equation}
is two-dimensional and defined by acceleration $a_x$ and steer velocity
$v_\delta$. The system dynamics are given as
\begin{align*}
	\frac{\mathrm{d}\bs{x}}{\mathrm{d}t} \hspace{-2pt} =\hspace{-2pt}
	\begin{bmatrix}
	v \cos(\psi+\beta)\\
	v \sin(\psi+\beta)\\
	\omega\\
	v_\delta\\
	\frac{1}{m}[F_{\hspace{-1pt}\text{fx}\hspace{-1pt}} \cos (\beta-\delta)\hspace{-1pt}+\hspace{-1pt}F_{\hspace{-1pt}\text{fy}\hspace{-1pt}} \sin (\beta-\delta)\hspace{-1pt}+\hspace{-1pt}F_{\hspace{-1pt}\text{rx}\hspace{-1pt}} \cos (\beta)\hspace{-1pt}+\hspace{-1pt}F_{\hspace{-1pt}\text{ry}\hspace{-1pt}} \sin (\beta)]\\
	\frac{1}{mv}[F_{\hspace{-1pt}\text{fy}\hspace{-1pt}} \cos (\beta-\delta) \hspace{-2pt}-\hspace{-2pt} F_{\hspace{-1pt}\text{fx}\hspace{-1pt}} \sin (\beta-\delta)\hspace{-2pt}-\hspace{-2pt}F_{\hspace{-1pt}\text{rx}\hspace{-1pt}} \sin (\beta)\hspace{-2pt}+\hspace{-2pt}F_{\hspace{-1pt}\text{ry}\hspace{-1pt}} \cos (\beta)]\hspace{-2pt}-\hspace{-2pt}\omega\\
	\frac{1}{I_\text{z}}\left([F_{\hspace{-1pt}\text{fx}\hspace{-1pt}}\sin(\delta) \hspace{-1pt}+\hspace{-1pt} F_{\hspace{-1pt}\text{fy}\hspace{-1pt}}\cos(\delta)]l_\text{f} \hspace{-1pt}-\hspace{-1pt} F_{\hspace{-1pt}\text{ry}\hspace{-1pt}}l_\text{r} \right)\\
	\frac{1}{I_\text{w}}(-r_w F_{\text{fx}}+t_{\text{b}}T_{\text{b}}+t_{\text{e}}T_{\text{e}})\\
	\frac{1}{I_\text{w}}(-r_w F_{\text{rx}}+(1-t_{\text{b}})T_{\text{b}}+(1-t_{\text{e}})T_{\text{e}})
	\end{bmatrix}
\end{align*}
where $r_\text{w}$ denotes the effective tire radius,  $I_{\text{w}}$ is the wheel inertia and $t_{\text{b}}$, $t_{\text{e}}$ are the split parameters between front and rear axle for the brake and engine torque, respectively. $m$ is the
vehicle mass, and $I_z$ the moment of inertia about the vertical axis of the vehicle center of gravity.
The lateral tire forces $F_{\cdot\text{y}}$ and longitudinal tire forces $F_{\cdot\text{x}}$ are computed using the Pacejka magic formula~\citep{Pacjeka2012}. The indexes r and f stand for rear and front tire, respectively. $T_{\text{e}}$ and $T_{\text{b}}$ are engine and brake torque, respectively. Both are computed from the longitudinal acceleration $a_{x}$.

The model is capable to compute lateral drift and omits conventional small angle approximations for steering and slip angles. 
Please consult the documentation of the Common Road library~\citep{CommonRoad} for a detailed description of the vehicle single track drift model. 
The model parameters of the single track drift model are given in Table~\ref{tab:parameters}.

\begin{table}
    \centering\small
    \caption{Model parameters of reference model (single track drift) and ODE
    model (single track model).}
    \bigskip
    \begin{tabular}{llcc}
        \toprule
        Name & Symbol & Unit & Value\\
        \midrule
        vehicle mass & $m$ & kg & 1225\\
        distance COG front axle & $l_f$ & m & 0.883\\
        distance COG rear axle & $l_r$ & m & 1.508\\
        friction coefficient & $\mu$ & - & 1.048\\
        cornering stiffness front and rear & $C_f$, $C_r$ &
        $\frac{1}{\mathrm{rad}}$ & 20.89\\
        center of gravity height & $h$ & m & 0.557\\
        moment of inertia about z axis & $I_z$ & $10^3\mathrm{kgm}^2 $ & 1.538\\
        moment of inertia wheels & $I_w$ & $10^3\mathrm{kgm}^2 $ & 1.700 \\
        split parameter for engine & $t_e$ & - & 1 \\
        split parameter for brake & $t_b$ & - & 0.76\\
        effective tire radius & $r_w$ & m & 0.344 \\
        \bottomrule
    \end{tabular}
    \label{tab:parameters}
\end{table}
\subsection{Training and validation data}\label{subsec:training-and-validation-data}

Three data samples were drawn by simulation of the single track drift reference
model in \texttt{Python} and imported in \texttt{Julia}. Each sample consists
of 100\,seconds simulation data on 0.1\,seconds sample rate. Each data sample
was split into training and validation set at $t=70\,\mathrm{s}$. The first
segment $0 < t < 70\,\mathrm{s}$ is the training set and the second segment
$70 \leq t < 100\,\mathrm{s}$ is the validation set in each of the tree data
samples.

\begin{figure}
    \centering
    \includegraphics[width=\columnwidth]{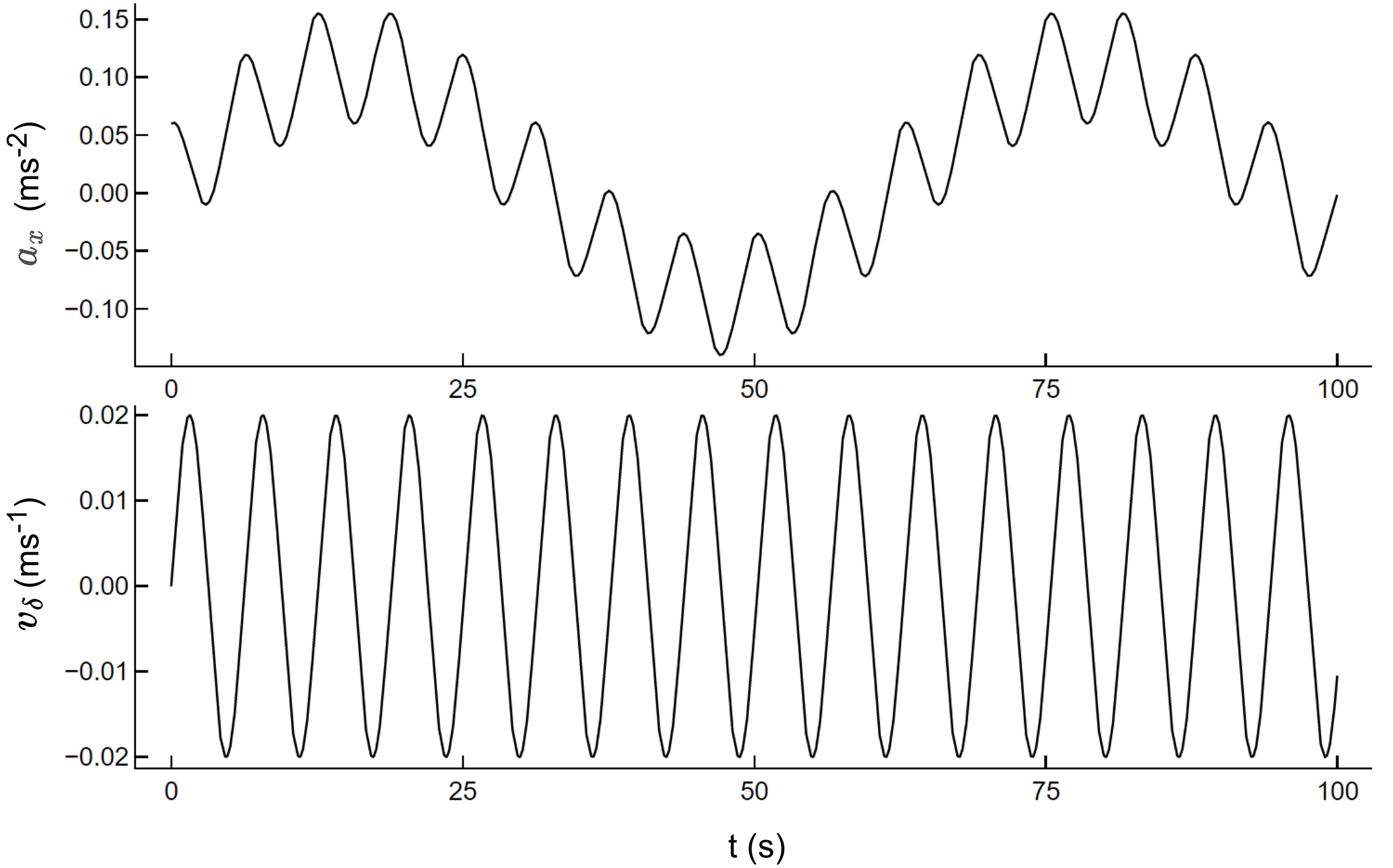}
    \caption{Exogenous input of acceleration ($a_x(t)$) and steer angle
    velocity ($v_\delta (t)$) of data sample three in top and bottom panel
    respectively. The x-axis denotes simulation time in seconds.}
    \label{fig:inputs}
\end{figure}

The first two data samples were used for a pre-training, while the last data
sample was used for calculation of training and validation error as well as for
generating the result plots for each model. The initial state vector at
$t=0\,\mathrm{s}$ and exogenous inputs of the third data sample are given in
the sequel, whereas the respective data for data sample one and two are
omitted for brevity. The initial state vector of the third data sample was
$\bs{x}_{t=0}=[0, 0, 0, 0, 25, 0, 0, 0, 0]^{\top}$, which means that only the
initial vehicle velocity was set to a non-zero value. The exogenous inputs of
data sample three became $a_x(t) = 0.01 + 0.05 \cos(t) + 0.1 \sin(0.1 t)$ and
$v_\delta (t) = 0.02 \sin(t)$ respectively. Figure~\ref{fig:inputs} presents
both inputs over simulation time.
Note that $\bs{x}_{t=0}$, $a_x(t)$, and $v_\delta (t)$ were slightly modified
in data sample one and two to enrich the coverage rate of the state space for
training of neural ODE and UDE model.

A z-score data scaler of the states and inputs was fit to the third data
sample. This scaler prepares the data for the neural network input layer and
for the cost function in the neural ODE and UDE model. In addition, the
scaler was used to normalize the reference states for adding white Gaussian
noise with zero mean and deviation $\sigma=0.025$ on each reference state.
Afterwards, the scaler transformed the noisy reference states back into
original space and the noisy reference states show comparable noise level.

\subsection{ODE model}\label{subsec:ode-model2}
The ODE model is inspired by the linear vehicle single track model, see, e.g., \citet{Althoff2017}. 
Its state vector consists of 
\begin{equation}
\bs{x} = [\Delta x, \Delta y, \psi, \delta, v, \beta, \omega]^{\top}
\label{eq:state}
\end{equation}
and the dynamics are given as
\begin{align}
\frac{\mathrm{d}\bs{x}}{\mathrm{d}t}  =
\begin{bmatrix}
v \cos(\psi+\beta)\\
v \sin(\psi+\beta)\\
\omega\\
v_\delta\\
a_{\text{x}}\\
\frac{1}{mv}[F_{\text{fy}}  +F_{\text{ry}}]-\omega\\
\frac{1}{I_\text{z}}\left( F_{\text{fy}}l_\text{f} - F_{\text{ry}}l_\text{r} \right)
\end{bmatrix}
\label{eq:single_track}
\end{align}
where the simplified lateral tire forces are 
\begin{align}
	F_{\text{fy}}\hspace{-2pt}=\hspace{-2pt}\mu C_{\text{f}} \frac{mgl_{\text{r}}}{l_{\text{r}}+l_{\text{f}}}\left(\delta - \frac{\dot{\psi}l_{\text{f}}}{v}-\beta\right), 
	F_{\text{ry}}\hspace{-2pt}=\hspace{-2pt}\mu
    C_{\text{r}}\frac{mgl_{\text{f}}}{l_{\text{r}}+l_{\text{f}}}\left(\frac{\dot{\psi}l_{\text{r}}}{v}-\beta\right),
\end{align}
where $\mu$ defines the tire road friction coefficient, $C_f,
C_r$ are the front and rear cornering stiffness, $h$ is the height of center
of gravity above road level, $g$ is the gravitational constant. Additional
simplifications compared to the single track drift model are that the wheel slip
and thus the wheel speeds as well as the longitudinal forces are neglected. Furthermore, a small angle approximation, i.e.~$\cos(x)\approx 1$ and $\sin(x)\approx x$, is carried out. The values of the vehicle parameters were identical in the single track drift and the single track model.

Especially the difference in the tire models cause a systematic model error
for the single track model (ODE model), compared with the single track drift model, which generated the training and validation data sets.
The single track model serves as ODE benchmark method in the sequel to show the benefit of using data in the following neural ODE and UDE methods.
\subsection{Neural ODE model}\label{subsec:neural-ode-model}
The neural ODE is given as
\begin{equation}
    \frac{\mathrm{d}\bs{x}}{\mathrm{d}t} = \mathrm{NN}(\mathcal{Z}\{[\bs{x},
    \bs{u}]^\top\}, \bs{\theta}, t),
    \label{eq:node}
\end{equation}
where $\mathcal{Z}\{\cdot\}$ means a z-score data transformation. The neural
ODE input layer is nine-dimensional and consists of z-score transformed
states and exogenous inputs. The hidden layer comprises a varying number of
fully connected neurons with $\tanh(\cdot)$ activation function and the
output layer models all seven derivatives of the states in~\eqref{eq:state}.
\subsection{UDE model}\label{subsec:ude-model}
The UDE model is given as
\begin{equation}
\frac{\mathrm{d}\bs{x}}{\mathrm{d}t} =
\begin{bmatrix}
v \cos(\psi + \beta)\\
v \sin(\psi + \beta)\\
\omega\\
v_{\delta}\\
\mathrm{NN}(\mathcal{Z}\{[\delta, v, \beta, \omega, a_x, v_\delta]^\top\},
\bs{\theta}, t)
\end{bmatrix},
\label{eq:ude_model}
\end{equation}
where the first four relations are identical with the ODE
model~\eqref{eq:single_track}.

Our aim is to remove simple and general physical relations from the learning
task of the NN in the UDE model. Hence, the kinematic relations of the four
first states are not part of the neural ODE network here. We assume that
these relations are in general correct and there is no need to apply machine
learning to reveal these certain equations. Accordingly, the neural network
within the UDE model is more compact compared with the neural network in the
neural ODE model. The neural network in the UDE has six inputs in the input
layer and one hidden layer of fully connected neurons with $\tanh(\cdot)$
activation function on varying size. The output layer models the derivatives
of the last three states of the state vector in~\eqref{eq:state}. This leads
to fewer network weights in the  UDE model.

Remark - While the ODE, neural ODE, and UDE model, share the same state
vector $\bs{x}$ and vector of exogenous inputs $\bs{u}$, the reference model
consists of two more states.

\subsection{Training}\label{subsec:training}

The ODE model does not require parameter estimation or training because it is
build from first principles and shares exact vehicle parameters with
reference model.

Neural ODE and UDE were deployed on varying size in the hidden layer to find
the optimal model complexity in accordance with Occam's razor. We varied the
hidden layer from five up to 12 neurons in neural ODE model and UDE model.
Table~\ref{tab:weights} lists the number of trainable network weights
dependent on the hidden layer size for the neural ODE and UDE model. Due to
smaller size in input and output layer, the UDE model has in general fewer
weights in the network weight matrix $\bs{\theta}$ compared with the neural
ODE network on same hidden layer size.

\begin{table}
    \centering\small
    \caption{Number of weights in neural network weight matrix $\bs{\theta}$
        over hidden layer size of neural ODE and UDE model.}
    \bigskip
    \begin{tabular}{lrrrrr}
        \toprule
        neurons in hidden layer & 5 & 8 & 10 & 12\\
        \midrule
        weights in neural ODE & 92 & 143 & 177 & 211\\
        weights in UDE & 53 & 83 & 103 & 123\\
        \bottomrule
    \end{tabular}
    \label{tab:weights}
\end{table}

The neural ODE model and UDE model were trained sequentially on the training
batches ($0<t<70\,\mathrm{s}$) of data sample one to three. In each training,
\texttt{Julia's} default random number generator initialized the neural
network weights to ensure reproducible results and this weight initialization
was repeated for five different random number seeds to explore the
sensitivity of the optimization result with respect to the initialization.
The number of neurons in the hidden layer controls the model complexity. Due
to the involved neural network, the neural ODE and UDE model fall under the
bias variance dilemma. To few neurons may lead to a model which cannot
represent the underlying process, which causes high bias in the model
estimates. Too many neurons may cause overfitting and poor prediction
capabilities, known as high variance. Hence, we varied the hidden layer size
to find the point where the training error and validation error are both
acceptable. Following Occam's razor principle, the best model is the model on
good accuracy and parsimony. On the other hand, poor initialization of the
neural network weights may cause poor optimization result because the
optimizer might get stuck in a local optima. We decided to use random weight
initialization for simplicity and spent the effort of five different
optimizations for each hidden layer size to reduce the risk of early inferior
local optima. This simple but effective setup worked in our experiments. The
random initialization of the weights did not result in numerical
problems due to initial stiff or unstable neural ODEs. Hence, more advanced
initialization methods like collocation pre-training, neutral pre-training,
or inclusion of model faders, as discussed in~\citep{Thummerer2022}, was not
required herein.

In total, we conducted twenty optimizations for neural ODE and UDE model
respectively (four different hidden layers times five initialization seeds). ADAM
algorithm~\citep{Kingma2014} optimized the neural network parameters on 2000
iterations for each data sample. The optimal learning rate for ADAM was found
by initial experiments. The hidden layer of neural ODE and UDE were adjusted
five to neurons (the smallest net in the sequel) and the learning rate
varied within $\alpha = [0.1, 0.075, 0.05,  0.025, 0.01, 0.001]$.

Standard training on the entire batch of training data failed in the
initial experiments, because the optimizer got stuck in local optima, which
resulted in constant horizontal state trajectories on high training and
validation error. The reason for the poor initial results are the
oscillations in the training data. Presenting all data at once with
oscillations caused ``flattened out'' trajectories and was recently reported
on other artificially and experimental data in~\citep{Turan2022}. There are
two solutions for this problem: incremental learning and multiple shooting.
In incremental learning, one splits the training data in smaller segments
and presents these segments iteratively to the optimizer. The first
training is done on segment one, the next training on segment one and two,
and so on. This approach is simple and robust but causes unwanted
computational load due to iteratively increasing data size and reuse of
training segments. In contrast, multiple shooting is more computational
efficient because the cost function is applied to individual data segments and
coupled through shooting variables to ensure smooth state transitions at the
training segments. Therefore, we applied multiple shooting function of
\texttt{DiffEqFlux.jl}~\citep{Rackauckas2019} in all experiments. The segment
size was adjusted to 80 data points and the continuity term was set to one.

\subsection{Performance criteria}\label{subsec:performance-criteria}

For neural ODE and UDE model, the training and validation error for data sample
three was evaluated with the sum of squared errors (SSE)
\begin{equation}
    \mathrm{SSE} = \sum\limits_{i=1}^n \left(\bs{x}_i -
    \bs{\widehat{x}_i}\right)^2,
    \label{eq:squarred_error}
\end{equation}
where $\bs{x}$ and $\widehat{\bs{x}}$ are matrices of reference and estimated
state trajectories at index $i$. To ensure equal influence of individual state
errors, the state trajectories were scaled with z-score transformation before
SSE computation. Note that the physical quantities range between $10^{-2}$ and
$10^2$ in the state space. Hence, certain states would be practically
neglected if the cost function would operate on untransformed data.

\section{Results and discussion}\label{sec:results_discussion}
First, we will present and discuss the benchmark result from ODE model. Then,
we will discuss the results of the initial training on varying learning rate
for neural ODE and UDE model and present the results of neural ODE and UDE
model with variation in hidden layer size and network weight initialization.
Finally, the best neural ODE and UDE model state trajectories are presented.

\subsection{ODE model}\label{subsec:ode-model}

\begin{figure}
    \centering
    \includegraphics[width=\columnwidth]{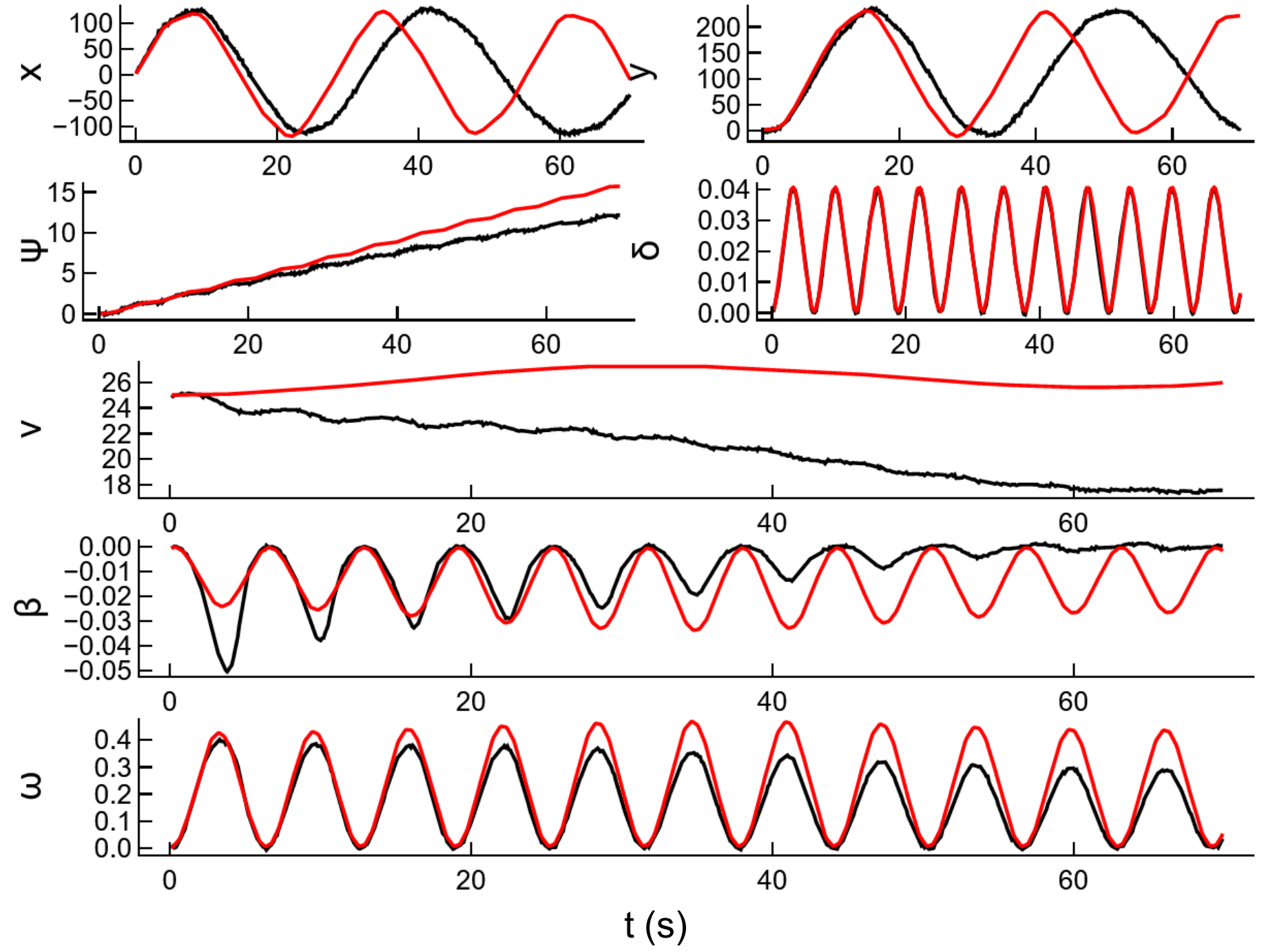}
    \caption{Estimated state trajectories of ODE model (red line) and
    reference model (black line) over the training batch. Note the large
    deviation between ODE mode and reference data in the  velocity $v$ (fifth
    panel).}
    \label{fig:ODE_states}
\end{figure}

Figure~\ref{fig:ODE_states} compares the simulated state trajectories of the
ODE single track model with the reference states from the single track drift
model. All states in Figure~\ref{fig:ODE_states} (and
Figures~\ref{fig:NODE_10_states},~\ref{fig:UDE_10_states},~\ref{fig:UDE_5_states})
are given in SI units over time in \si{\second}. The positions $x$ and $y$
in \si{\meter}, yaw angle $\psi$ and steer angle $\delta$ in \si{\radian},
velocity $v$ in \si{\meter\per\second}, slip angle $\beta$ in \si{\radian}, and
yaw rate $\omega$ in \si{\radian\per\second}.

Although all model parameters like vehicle mass, cornering stiffness
are identical in ODE in reference model, the ODE model shows poor accuracy.
The positions in panel one and two, yaw angle $\psi$ in panel three, velocity
$v$ in panel five drift away from reference data in time. The ODE
model overestimates the oscillation in the slip angle $\beta$ in panel six
and in the yaw rate $\omega$ in the last panel. To sum up, the neglected
drift effect in the simple ODE model causes a large model error in
comparison with the reference data from the single track drift model. Due to
its poor accuracy, we will exclude the ODE model in the following in detail
result presentation.

\subsection{Varying the learning rate for neural ODE and UDE model}
\label{subsec:results_learning_rate}

Table~\ref{tab:learning_rate} lists the training and validation error of
neural ODE and UDE model for each learning rate. The smallest training error
of neural ODE was found by a learning rate of 0.05, whereas the UDE model
gave the smallest training error at $\alpha=0.025$. Overall, the UDE training
and validation errors are less sensitive to variation in learning rate than
the neural ODE errors. Specifically, the smallest and largest learning rates
cause poor accuracy in the neural ODE model, where the UDE model still
produces acceptable results. Finally, we selected $\alpha=0.05$ for ADAM
optimization of neural ODE weights and $\alpha=0.025$ for UDE weights.

\begin{table}
    \centering\small
    \caption{Training and validation error over learning rate for neural ODE
        (nODE) and UDE model, each with hidden layer of five neurons. The
        errors were computed on z-score transformed data with sum of squared
        errors. The smallest training errors are written in bold. The sum of squared error of the ODE model for simulation scenario used in the training is 7988, yielding a performance improvement of $7988/125 \approx 63\%$ and $7988/117 \approx 68\%$ for the nODE and UDE model, respectively.}
    \bigskip
    \begin{tabular}{lrrrr}
        \toprule
        & \multicolumn{2}{c}{training error} & \multicolumn{2}{c}{validation
        error}\\
        learning rate & nODE & UDE          & nODE     & UDE\\
        \midrule
        0.1   & 2327   & 122          & 1301  & 50\\
        0.075 & 230    & 249          & 539   & 225\\
        0.05  & \textbf{125} & 134    & 972   & 125\\
        0.025 & 143    & \textbf{117} & 810   & 47\\
        0.01  & 104202 & 145          & 895   & 103\\
        0.001 & 128670 & 163          & 10434 & 31\\
        \bottomrule
    \end{tabular}
    \label{tab:learning_rate}
\end{table}

\subsection{Varying hidden layer and initialization in neural ODE and UDE
model}\label{subsec:varying-the-hidden-layer-size-in-neural-ode-and-ude-model}

Table~\ref{tab:hidden_layer} shows the training and validation error over
increasing number of neurons in the hidden layer, and on varying random
initialization for the neural ODE and UDE model. In contrast to the learning
rate selection, we focus here on small validation error to find the best
hidden layer size and initialization for neural ODE and UDE
model. Accordingly, the best neural ODE and UDE result was found with ten
neurons in the hidden layer. However, the best initialization for neural ODE
was the fifth, whereas best UDE model was optimized on first initialization.
Therefore, the in depth results of neural ODE and UDE model will be presented
with 10 in the hidden layer and the respective random number
setting in the next sections. Strikingly, the optimization is highly
sensitive to the initialization of network. Some initialization cause large
validation error. For instance the UDE model with twelve neurons and random
number seed two appears as outlier. Therefore, either repetitive
initializations on different seeds (as done herein), or more advanced
initializations methods must be applied in neural ODE experiments in order to
compare different model structures.

Again, the neural ODE model shows often larger variation in errors than the
UDE model and in the error level of neural ODE model is roughly one order of
magnitude larger than the validation error of the UDE model. However, both
models show some outliers in the validation when the initialization caused a
remarkably poor fit. In neural ODE, a steady decrease of training error over
growing number of neurons in the hidden layer is visible, which is supported
by machine learning theory. The evolution of training error in UDE model is
rather stable, which does not result in concrete conclusions.

\begin{table}
    \centering\small
    \caption{Training and validation error of neural ODE (nODE) and UDE model
    over varying hidden layer size. Rng colum denotes experiments where
    the random number seed of network weights initialization was varied. The
    smallest validation errors are written in bold.}
    \bigskip
    \begin{tabular}{rrrrrr}
        \toprule
        && \multicolumn{2}{c}{training error} & \multicolumn{2}{c}{validation
        error}\\
        \parbox[b]{4em}{hidden layer size} & rng & nODE & UDE & nODE & UDE\\
        \midrule
        \multirow{5}{*}{5} & 1 & 125  & 117 & 972   & 47\\
                           & 2 & 216  & 126 & 863   & 120\\
                           & 3 & 350  & 82  & 3667  & 156\\
                           & 4 & 1926 & 148 & 1185  & 53\\
                           & 5 & 247  & 118 & 1209  & 24\\
        \cline{2-6}
        \multirow{5}{*}{8}  & 1 & 49  & 127 & 821  & 59\\
                            & 2 & 27  & 115 & 681  & 39\\
                            & 3 & 59  & 189 & 1045 & 156\\
                            & 4 & 20  & 182 & 679  & 124\\
                            & 5 & 29  & 133 & 851  & 22\\
        \cline{2-6}
        \multirow{5}{*}{10} & 1 & 20  & 119 & 1343 & \textbf{16}\\
                            & 2 & 16  & 116 & 942  & 41 \\
                            & 3 & 15  & 152 & 816  & 133 \\
                            & 4 & 72  & 143 & 1159 & 413 \\
                            & 5 & 13  & 206 & \textbf{349} & 343 \\
        \cline{2-6}
        \multirow{5}{*}{12} & 1 & 20  & 173  & 614  & 67 \\
                            & 2 & 25  & 199  & 547  & 4686 \\
                            & 3 & 19  & 163  & 623  & 260 \\
                            & 4 & 22  & 140  & 759  & 98 \\
                            & 5 & 16  & 125  & 440  & 117 \\
        \bottomrule
    \end{tabular}
    \label{tab:hidden_layer}
\end{table}

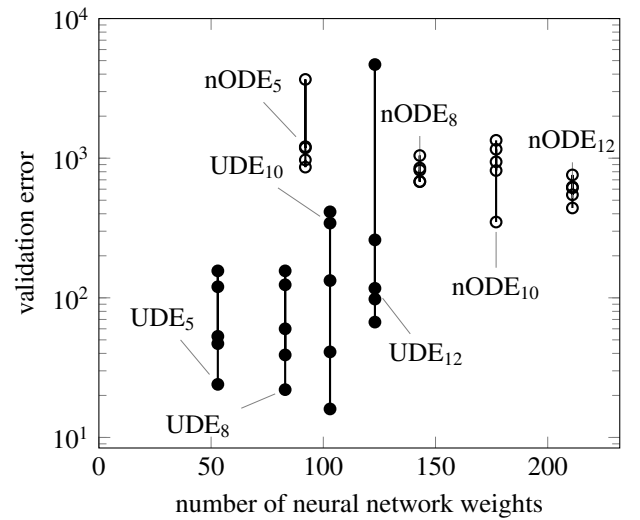
\begin{figure}
    \centering
\pgfplotsset{
  every axis plot/.append style={thick},
    every mark/.append style={draw=black}
}
\begin{tikzpicture}
\begin{axis}[ymode=log, xlabel=number of neural network weights,
    ylabel=validation error, ymax=1e4, xmin=0]

    \addplot [mark=o] coordinates {
        (92,972.3) (92,863) (92,3667) (92,1185) (92,1209)}
        node[pin=105:{nODE\textsubscript{5}}]{};

    \addplot [mark=o] coordinates {
        (143,821) (143,681) (143,1045) (143,851) (143,679)}
        node[pin=90:{nODE\textsubscript{8}}]{};

    \addplot [mark=o] coordinates {
        (177,1343) (177,942) (177,816) (177,1159) (177,349)}
        node[pin=270:{nODE\textsubscript{10}}]{};

    \addplot [mark=o] coordinates {
        (211,614) (211,547) (211,623) (211,759) (211,440)}
        node[pin=90:{nODE\textsubscript{12}}]{};

    \addplot [mark=*] coordinates {
        (53,47) (53,120) (53,156) (53,53) (53,24)}
        node[pin=105:{UDE\textsubscript{5}}]{};

    \addplot [mark=*] coordinates {
        (83,60) (83,39) (83,156) (83,124) (83,22)}
        node[pin=200:{UDE\textsubscript{8}}]{};

    \addplot [mark=*] coordinates {
        (103,16) (103,41) (103,133) (103,413) (103,343)}
        node[pin=135:{UDE\textsubscript{10}}]{};

    \addplot [mark=*] coordinates {
        (123,67) (123,4686) (123,260) (123,98) (123,117)}
        node[pin=275:{UDE\textsubscript{12}}]{};

\end{axis}
\end{tikzpicture}    
    \caption{Validation error of all neural ODE (nODE) and UDE model over the
    number of neural network weights. The subscripts in the model names
    denote the hidden layer size. For instance, UDE\textsubscript{10}
    means the UDE model with ten neurons in the hidden layer. Dots
    connected with vertical lines denote repeated trainings on different
    initialization of the network weights. Please note the logarithmic scale of
    the validation error.}
    \label{fig:model_selection}
\end{figure}

Figure~\ref{fig:model_selection} presents the same data as
Table~\ref{tab:hidden_layer} in another format. Here, the validation errors
for each neural ODE and UDE model is drawn over the number of neural network
weights, which represents the model complexity. Models on same structure but
different initializations are connected with solid vertical lines. Apart from
one outlier on twelve neurons, all UDE model build a cluster in the lower
left edge of the figure, which means that almost all UDE models show higher
accuracy than the neural ODE models. Moreover, we found initializations where
the smallest UDE models on five and eight neurons were roughly as accurate as
the best UDE model with ten neurons. Hence, one could also argue to prefer a
smaller UDE model in practical applications than the model that we have
selected. Add to this, the complexity of many UDE models is smaller than the
neural ODE models. Hence, the UDE models are in general the better choice
following Occam's razor. With focus on accuracy, neural ODE and UDE with 10
neurons in hidden layer are best in each model class.

\subsection{Neural ODE\textsubscript{10} model}\label{subsec:neural-10-model}

\begin{figure}
    \centering
    \includegraphics[width=\columnwidth]{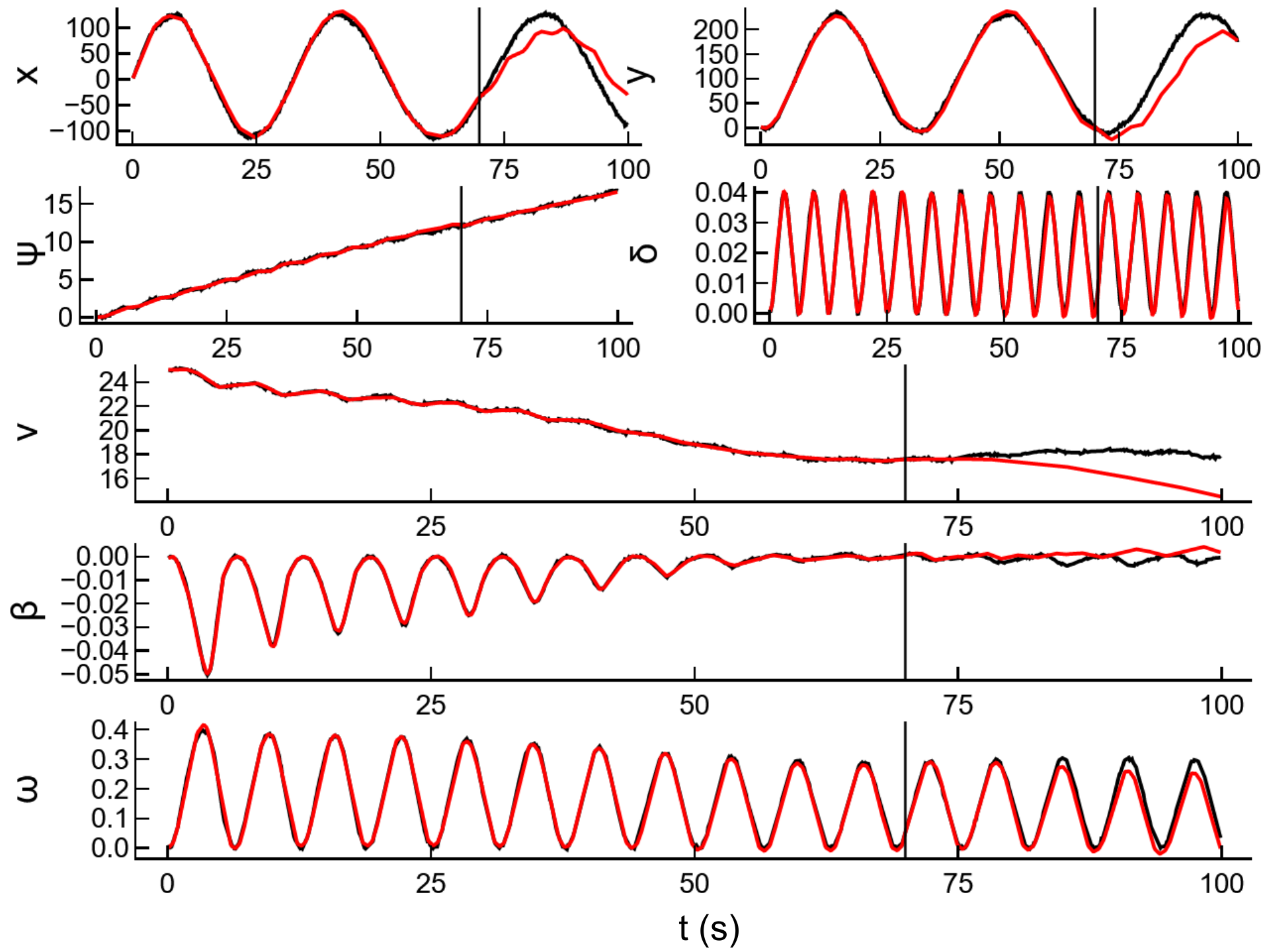}
    \caption{Estimated state trajectories of neural ODE\textsubscript{10}
    model with 10 neurons in hidden layer (red line) and reference model
    (black line). The vertical line at $t=70\,\mathrm{s}$ marks the split
    between training and validation data.}
    \label{fig:NODE_10_states}
\end{figure}

Figure~\ref{fig:NODE_10_states} gives the estimated states over simulation
time for the neural ODE model with 10 neurons in the hidden layer on rng seed
five. The model shows high accuracy for the training batch ($t<70\,
\mathrm{s}$), but the states diverge from reference data in the validation
set from $t>70\,\mathrm{s}$ onwards. This accuracy drop in the validation is
mainly driven by the position states in first and second panel, yaw angle in
third panel, and velocity in fifth panel. The other states show slight error
in validation. Compared with the benchmark ODE model in
Figure~\ref{fig:ODE_states}, the neural ODE model is more accurate. The
numerical values for training and validation error of the neural
ODE\textsubscript{10} model can be found in Table~\ref{tab:hidden_layer}.

\subsection{UDE\textsubscript{10} model}\label{subsec:ude-10-model}

\begin{figure}
    \centering
    \includegraphics[width=\columnwidth]{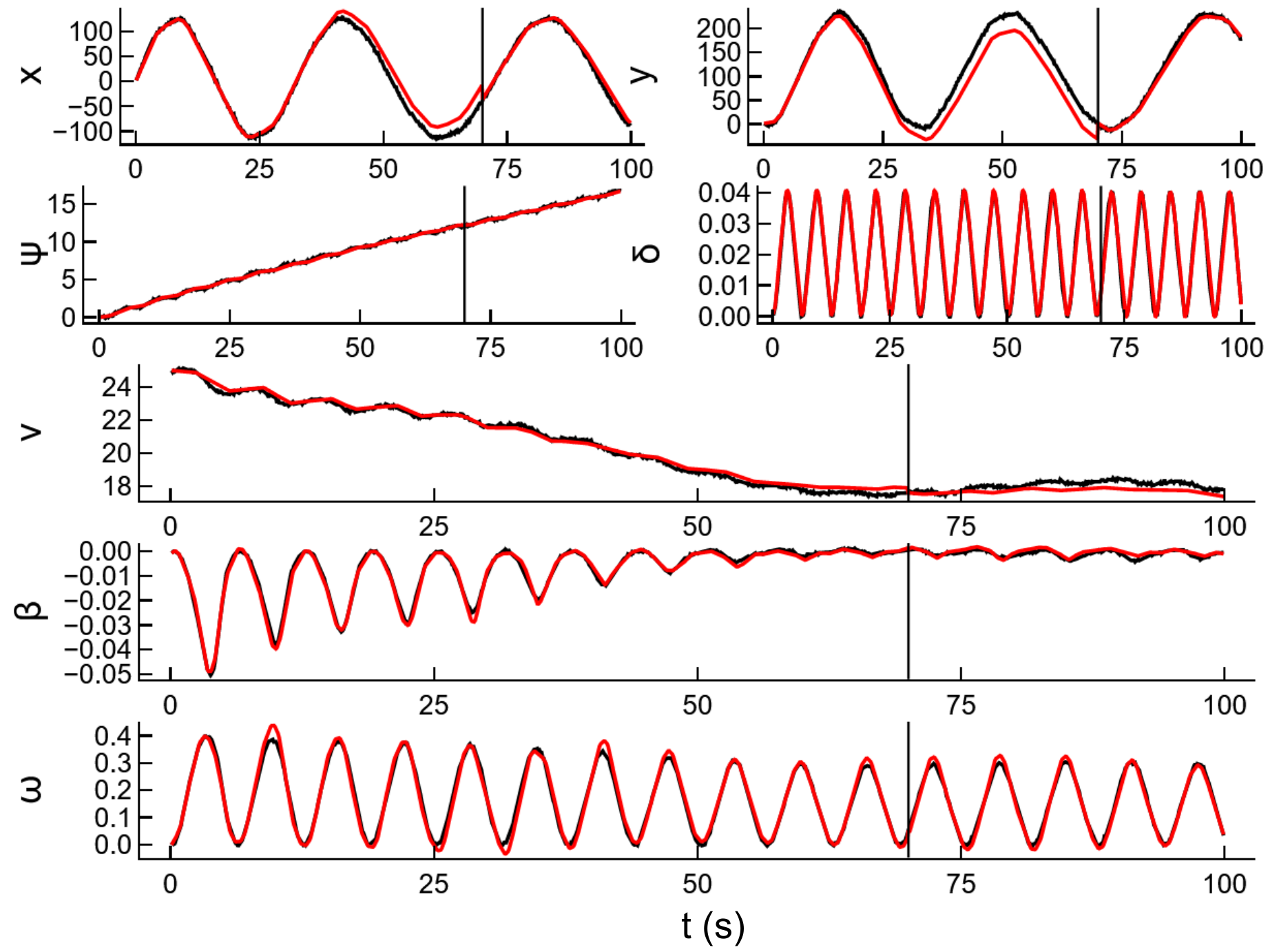}
    \caption{Estimated state trajectories of UDE\textsubscript{10} model with
    10 neurons in hidden layer (red line) and reference model (black line).
    Training and validation set is split at $t=70\,\mathrm{s}$, indicated by
    the black vertical line.}
    \label{fig:UDE_10_states}
\end{figure}

Figure~\ref{fig:UDE_10_states} gives the estimated states over
simulation time for the UDE model with 10 neurons in the hidden layer on rng
seed one. UDE\textsubscript{10} is not as accurate as neural
ODE\textsubscript{10} in training batch, but superior accurate in the
validation set among all models. Considering the relative small number neural
network weights of UDE\textsubscript{10} (103\,weights), this model is the
best model in the set of candidate models. The smaller accuracy in training
but higher accuracy invalidation indicates that the model generalized the
data better than neural ODE\textsubscript{10} model. However, the first two
position states of UDE\textsubscript{10} start to diverge from reference
roughly in the middle of the training batch at $t>30\,\mathrm{s}$. On the
other hand, the introduced physical relations for the yaw angle $\mathrm{d}\psi/
\mathrm{d}t= \omega$ and steer angle $\mathrm{d}\delta/ \mathrm{d}t=
v_\delta$ in~\eqref{eq:ude_model} appear to improve the accuracy compared
with neural ODE\textsubscript{10} model.

\subsection{UDE\textsubscript{5} model}\label{subsec:ude-5-model}

\begin{figure}
    \centering
    \includegraphics[width=\columnwidth]{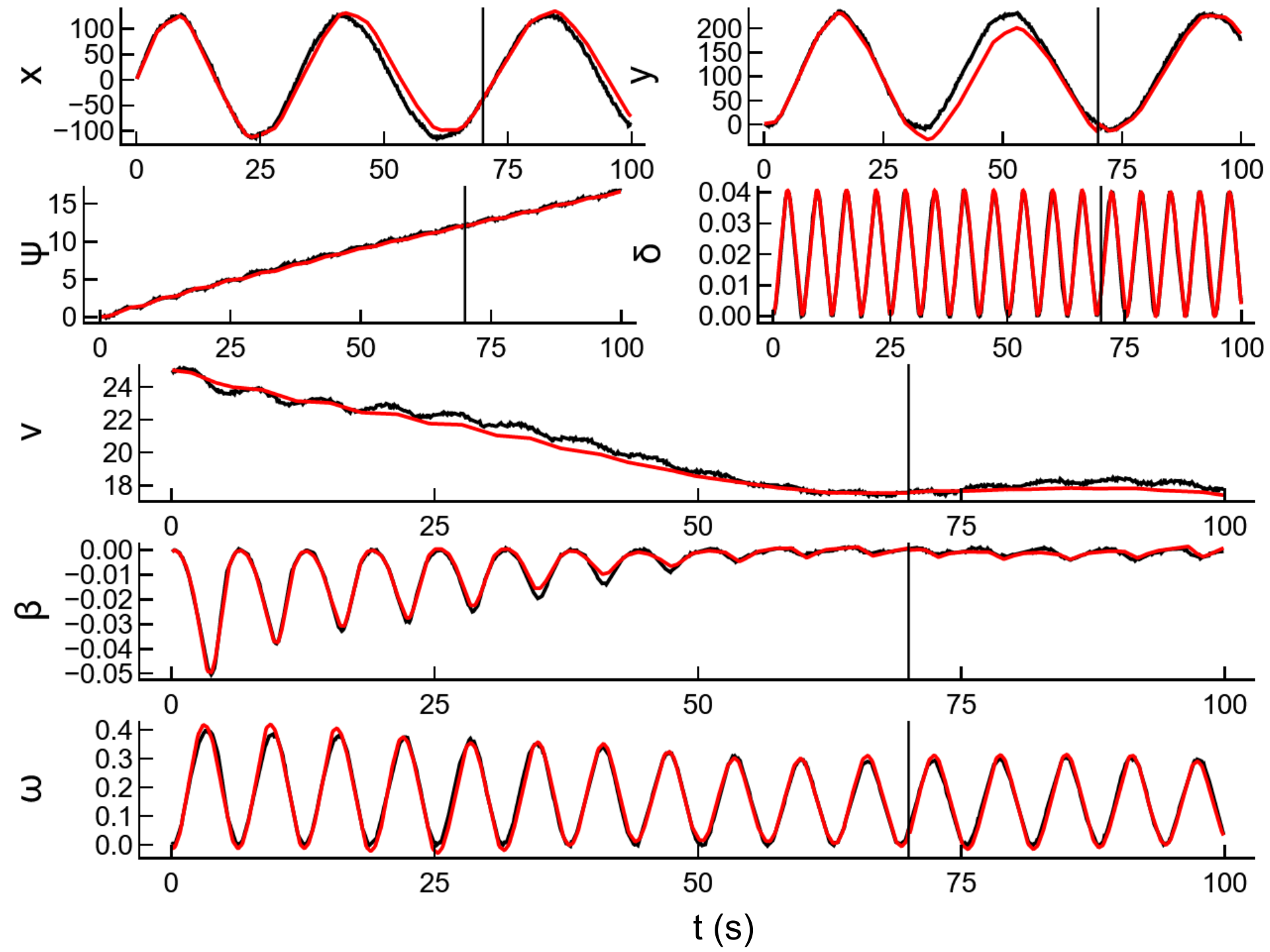}
    \caption{Estimated state trajectories of UDE\textsubscript{5} model with
    5 neurons in hidden layer (red line) and reference model (black line).
    Training and validation set is split at $t=70\,\mathrm{s}$, indicated by
    the black vertical line.}
    \label{fig:UDE_5_states}
\end{figure}

Finally, Figure~\ref{fig:UDE_5_states} gives the estimated states over
simulation time for the UDE model with 5 neurons in the hidden layer on rng
seed five. This model is nearly as accurate as the previous
UDE\textsubscript{10} model, but more efficient in required storage for
network weights, in training, and in evaluation time. Therefore, the compact
UDE\textsubscript{5} might be preferable in embedded functions where storage
and processing power are very limited. However, the UDE\textsubscript{5}
model does not capture all dynamics in the velocity between $20 < t < 50$
seconds compared with UDE\textsubscript{10} and reference data. Instead, the
velocity trajectory is rather flat and smooth.

\section{Conclusions}\label{sec:conclusions}

This comparative study of different modeling methods, ranging from
physics based modeling with a vehicle single track ODE model, over black box
modeling with neural differential equations (NODE), to hybrid modeling through
universal differential equations (UDE), showed that the UDE approach
was superior in terms of validation accuracy and model complexity.
In specific, the UDE\textsubscript{10} model corrected missing
physical effects of a conventional vehicle single track ODE model through the
neural ODE part in the states of vehicle velocity, slip angle and yaw rate.
Add to this, the hybrid model took favor of physics based kinematic state
equations for vehicle positions, yaw angle, and steer angle in order to
improve the prediction accuracy for these states compared with the pure black
box neural ODE model.

Moreover, the combination of physics based differential equations and neural
differential equations allowed to reduce the size of the neural network
significantly compared with a pure black box model, which helps to reduce
training and evaluation time and limits required storage for network weights
in embedded applications. In conclusion, the UDE\textsubscript{10}
combines to a certain extent the predictability, generalization and
interpretability of physical modeling with adaptation to training data of
machine learning and was superior accurate than each modeling approach
individually.

In the future, more investigation is needed to study how the hybrid model
performs in other scenarios. We are looking also for methods to add
uncertainty estimates to the modeled states to receive a
confidence measure. This is especially important for states modeled by a 
neural network and allows for a credibility assessment of the model.
Another field of research is to detect and prevent the evaluation of the data-based
part far beyond the dynamics captured by the seen learning data. In this case one 
would rely on extrapolation capabilities, which a data-based model does not have in 
general. These measures are required for accreditation of hybrid models in real 
world applications such as advanced driver assistance systems.

\section*{Declaration of interests}
This work was supported by the ITEA3-Project UPSIM (Unleash Potentials in
Simulation) \textnumero\,19006, see: https://www.upsim-project.eu/ for more
information. Add to this, all authors report financial support, provided by
Robert Bosch GmbH and Stephan Rhode has patent pending to German Patent and
Trademark Office.

\section*{CRediT authorship contribution statement}
\textbf{Stephan Rhode:} Conceptualization, Methodology, Software, Validation,
Investigation, Writing - Original Draft, Writing - Review \& Editing,
Visualization. \textbf{Fabian Jarmolowitz:} Writing - Original Draft.
\textbf{Felix Berkel:} Writing - Original Draft.


\bibliographystyle{elsarticle-harv}
\bibliography{references}

\end{document}